\documentclass[reprint,aps,prb,floatfix,showpacs]{revtex4-1}
\usepackage{epsf}
\usepackage{epsfig}
\usepackage{float}
\usepackage{graphicx}
\usepackage{epstopdf}
\usepackage{amssymb}
\usepackage{amsmath}
\usepackage{comment}
\usepackage{color}
\usepackage{setspace}
\usepackage{comment}

\def\k{{{\bf k}}}
\def\om{{\omega}}

\def\g{{\bf{g}}}

\def\be{\begin{equation}}
\def\ee{\end{equation}}
\def\ber{\begin{eqnarray}}
\def\eer{\end{eqnarray}}

\def\g0{{\gamma_0}}

\def\Im{{\mbox{Im}}}

\def\prb{Phys.\ Rev.\ B }
\def\prl{Phys.\ Rev.\ Lett.\ }
\def\jpcm{J.\ Phys.\ Condens.\ Matter }

\def\rmp{Rev.\ Mod.\ Phys.\ }

\def\EPL{ Europhys.\  Lett.\ }

\newcommand\bear{\begin{eqnarray}}
\newcommand\eear{\end{eqnarray}}
\newcommand\bea{\begin{align}}
\newcommand\ena{\end{align}}

\begin{document}
\title{Interaction-Induced Topological and Magnetic Phases \\ in the Hofstadter-Hubbard Model}
\author{Pramod Kumar}
\author{Thomas Mertz}
\author{Walter Hofstetter}
\affiliation{Institut f{\"u}r Theoretische Physik, Johann Wolfgang Goethe-Universit{\"a}t, 60438 Frankfurt/Main, Germany}

\begin{abstract}
 Interaction effects have been a subject of contemporary interest in topological phases of matter. But in the presence of interactions, the accurate determination of topological invariants in their general form is difficult due to their dependence on multiple integrals containing Green's functions and their derivatives. Here we employ the recently proposed ``effective topological Hamiltonian" approach to explore interaction-induced topological phases in the time-reversal-invariant Hofstadter-Hubbard model. Within this approach, the zero-frequency part of the self-energy is sufficient to determine the correct topological invariant. We combine the topological Hamiltonian approach with the local self-energy approximation, both for the static and the full dynamical self-energy evaluated using dynamical mean field theory (DMFT), and present the resulting phase diagram in the presence of many-body interactions. We investigate the emergence of quantum spin Hall (QSH) states for different interaction strengths by calculating the $\mathbb{Z}_2$ invariant. The interplay of strong correlations and a staggered potential also induces magnetic long-range order with an associated first order transition. We present results for the staggered magnetisation ($m_s$), staggered occupancy ($n_s$) and double occupancy across the transition.

\end{abstract}

\pacs{topological Hamiltonian, cold atoms}

\maketitle

\section{Introduction}
\label{sec:intro}
In recent times, topological states have gained considerable attention, as they open an avenue towards studying novel phases of matter such as the quantum spin Hall (QSH) state, topological insulators and topological superconductors~\cite{Hasan10}. The QSH state can be considered as two decoupled copies of an integer quantum Hall  insulator with opposite chirality for the two different spin species, as proposed by Kane and Mele~\cite{Kane05}. The QSH state is insulating, with a topological band structure and a spin Hall current related to gapless edge states, which are protected by time reversal symmetry.
Achieving opposite chirality for two different species is not feasible with ``real'' electromagnetic fields. However, it is possible in ultracold atom systems using synthetic gauge fields~\cite{Dalibard11, Lin11}. Recently the non-interacting time-reversal-invariant (TRI) Hofstadter model has been realised experimentally in ultracold atoms using laser assisted tunnelling in tilted optical lattices~\cite{Aidelsburger13}.
The simplest way to theoretically include correlation effects in the QSH insulator is to add a local Hubbard interaction term to a model with spin-orbit coupling. Several interacting model Hamiltonians have been proposed, such as the Kane-Mele Hubbard model~\cite{Rachel10}, the Bernevig-Hughes-Zhang (BHZ) Hubbard model~\cite{Amaricci15, Roy16}, and the TRI Hofstadter Hubbard (HH) model~\cite{Goldman10,Cocks12}. Interacting model Hamiltonians which break time-reversal symmetry such as the Haldane-Hubbard model~\cite{Varney11} and the Kondo lattice model~\cite{Werner13} have been studied as well in the context of topological quantum phase transitions. 

    A topological invariant can be defined for QSH insulators in terms of a $\mathbb{Z}_2$ invariant which is $0$ for trivial insulators and $1$ for QSH insulators~\cite{Kane05,Hasan10}.
Several analytical and numerical methods exist for calculating these topological invariants for the non-interacting case~\cite{Hohenadler13}. 
Since time-reversal symmetry protects topological states against weak interactions, the bulk-boundary correspondence can, in principle, be used to determine topological transitions for finite interactions~\cite{Wang112, Wu12}. 
This approach suffers from difficulties such as carrying out numerical calculations for large enough system with open boundary conditions to identify the gapless edge states. 
Correlation effects and tunnelling  between edges can gap out the  edge states of the topologically non-trivial phase. 
In recent work~\cite{Cocks12, Orth13}, the bulk-boundary correspondence has been exploited to extract the topological invariant from the real-frequency spectral function calculated by dynamical mean field theory (DMFT). The extraction of the real-frequency spectrum was carried out by analytic continuation using the maximum entropy method.  
This is numerically demanding and can possibly include statistical as well as numerical errors. 
 The topological invariant can be directly extracted from the quantized Hall conductivity, although this is limited to the spin-conserving case~\cite{Yoshida12}. Moreover, calculating the Hall conductivity in the presence of interactions can  itself be difficult. Recently, a generalised topological invariant, termed as spin-Chern number has been proposed using the many-body state to explore disorder and spin-mixing perturbations~\cite{Xiao10,Sheng06}. Moreover it has also been discussed for the non-interacting Kane-Mele~\cite{Sheng06} and Hofstadter model~\cite{Goldman10}. Although this provides a general method to calculate the $\mathbb{Z}_2$ invariant, its numerical evaluation can be technically infeasible in the context of numerical methods like quantum Monte Carlo (QMC).
    Another very important development in the calculation of the Chern number or $\mathbb{Z}_2$ invariant was to represent them in terms of the full interacting single-particle Green's function~\cite{Ishikawa86, Wang10, Gaurarie11}. This relation provides insight into the bulk-boundary correspondence and interaction effects by analysing the Green's function only. However, carrying out the required multiple integrals of derivatives of the Green's function can be numerically challenging. 
A further simplification has therefore been proposed~\cite{Wang12} which is termed ``topological Hamiltonian". In this approach, assuming an adiabatic connection to the non-interacting state, the Green's function at zero frequency is sufficient to determine the topological invariant. For practical purposes, the topological Hamiltonian  can be used for calculating the Chern number or the $\mathbb{Z}_2$ invariant with methods that are applicable to non-interacting systems. In this work, we use the topological Hamiltonian approach combined with a local self-energy within Hartree-Fock (HF) and DMFT to calculate the effective non-interacting Hamiltonian. Further we employ the method by Fukui and Hatsugai~\cite{Fukui07} to calculate the $\mathbb{Z}_2$ invariant using this effective Hamiltonian and evaluate the effect of the dynamical contribution of the many-body interaction on the QSH state. The topological Hamiltonian combined with DMFT has been used in several recent works in the context of correlation-induced topological phases~\cite{Amaricci15, Werner13, Zdulski15,Vanhala15}. 
    
    The TRI HH model studied here is also related to the $2$D ionic Hubbard model on a square lattice, where phases such as band insulator (BI), metal and Mott insulator (MI) were reported~\cite{Paris07,Bouadim07}. Moreover in previous work~\cite{Cocks12,Orth13} magnetic quantum phase transitions from Ne{\`e}l- to spiral and collinear ordered states have been reported for the TRI HH model. Here we specifically choose zero spin-orbit coupling (where spin is conserved) where the magnetic order is Ne{\`e}l type and explore the nature of the magnetic transition with an applied staggering potential.
    
    The paper is structured as follows. We first introduce the
TRI HH model, followed by the formalism of the topological Hamiltonian with local self-energy approximation,
which is used to incorporate the effects of correlations. In section III A, we present the phase diagram in the presence of interactions which is the central result of this work. In sections III B and III C, we discuss the behavior of the self-energy and the resulting magnetic transitions respectively. In sections III D, we compare predictions for topological bands based on the topological Hamiltonian and on the bulk-boundary correspondence.  In section III E, we present the phase diagram for  zero spin-orbit coupling. In section III F, we demonstrate the magnetic transition for finite staggering field. In the final section III G, we show the staggered occupancy and double occupancy for finite staggering field and comment on related aspects of the $2$D ionic Hubbard model.
\section{Model and formalism}
\label{sec:model}
\subsection{Time-reversal-invariant Hofstadter-Hubbard model}
We explore interaction effects in the TRI HH model~\cite{Goldman10,Cocks12}.
The Hamiltonian for the Hofstadter-Hubbard model is expressed in standard second-quantization notation as
\begin{align}
H=&-\sum_j\Big[t_x c_{j+\hat{x}}^\dagger e^{-i2\pi\gamma\sigma^x}c_j+t_y c_{j+\hat{y}}^\dagger e^{i2\pi\alpha x\sigma^z}c_j 
+ \text{h.c.} \Big] \\ \nonumber 
&+\sum_j \left[(-1)^x\lambda_x \ c_j^\dagger c_j-\mu \ c_j^\dagger c_j +U n_{j\uparrow}n_{j\downarrow}\right]\label{eq1},
\end{align}
where $c_j^\dagger=(c_{j\uparrow}^\dagger,c_{j\downarrow}^\dagger)$ is the creation operator at lattice site $j = (x, y)$, $\sigma^{x(z)}$ are  Pauli
matrices and $\hat{x} = (1, 0)$, $\hat{y} = (0, 1)$ are unit vectors. The first term is a Rashba-type spin-orbit coupling  term of strength $\gamma$, which determines the coupling between the two spin species via $e^{i2\pi\gamma \sigma^x}=\cos(2\pi\gamma)\mathbf{1}+i\sigma^x\sin(2\pi\gamma)$. In the second term, the value of $\alpha$ determines the dimensionless strength of the artificial flux. The third term is a staggered potential $\lambda_x$ in $x$-direction, which defines a sub-lattice structure. $\mu$ is the chemical potential and the last term is the on-site Hubbard interaction of strength $U$. We consider isotropic hopping where $t_x=t_y=t$. It is well known that the TRI Hofstadter model with $U=0$ has a rich phase diagram including semi-metal (SM), quantum spin Hall (QSH) state and band insulator (BI)~\cite{Goldman10}. 

  In order to explore the  effect of interactions on topological states, we use the topological Hamiltonian approach and combine it with local self-energy approximations, for both the static and the full dynamical self-energy evaluated using DMFT. In this work, we study the system in the homogeneous case, i.e. periodic in both $x$- and $y$-directions. To evaluate the edge states, we study the system in the stripe geometry with open-periodic boundary condition, i.e. open in $x$-direction and periodic in $y$-direction.  In the following, we present the details of our method.   
\subsection{Topological Hamiltonian and dynamical mean field theory}
\label{subsec:formalism}
In the presence of interactions, the Chern number or the $\mathbb{Z}_2$ invariant can be formulated in terms of the single particle Green's function~\cite{Ishikawa86,Wang10,Gaurarie11}. The accurate numerical determination of the topological invariant is demanding due to multiple integrals containing Green's functions and their derivatives. 
It is an equally formidable task to accurately determine the frequency- and momentum-resolved Green's function for interacting systems~\cite{Hohenadler13}. Therefore, in order to calculate the topological invariant in the presence of interactions, we employ the recently proposed topological Hamiltonian approach~\cite{Wang12}. The topological Hamiltonian in the presence of interactions can be written as
\begin{eqnarray}
\hat{h}_t=-\hat{G}^{-1}_{ij}(i\omega_n=0)=\hat{h}_{ij}+\hat{\Sigma}_{ij}(i\omega_n\rightarrow0)\label{eq3},
\end{eqnarray}
where $\hat{h}_{ij}$ is the non-interacting part of the Hamiltonian and $\hat{\Sigma}_{ij}(i\omega_n)$ is the self-energy. Further, we combine the topological Hamiltonian approach with the local self-energy approximation, i.e. $\hat{\Sigma}_{ij}(i\omega)\equiv \hat{\Sigma}_{ii}(i\omega_n)\delta_{ij}$, which leads to
\begin{eqnarray}
\hat{h}_t=-\hat{G}^{-1}_{ij}(i\omega_n=0)=\hat{h}_{ij}+\hat{\Sigma}_{ii}(i\omega_n \rightarrow 0)\delta_{ij} \label{eq4}.
\end{eqnarray}
In the Fermi-liquid regime, where the self-energy is analytical, its imaginary part has a linear dependence at small frequencies, i.e.  $\text{Im}\Sigma_{ii}(i\omega_n)\sim i\omega_n$, and thus tends to zero for $i\omega_n\rightarrow 0$. Therefore the above equation can be written as
\begin{eqnarray}
\hat{h}_t=-\hat{G}^{-1}_{ij}(i\omega_n=0)=\hat{h}_{ij}+\hat{\Sigma}_{ii; R}(i\omega_n \rightarrow 0)\delta_{ij}, \label{eq5}
\end{eqnarray}
where $\hat{\Sigma}_{ii; R}(i\omega_n)$ is the real part of the local self-energy. The general form of the local self-energy can be  written as~\cite{Antoine96}
\begin{equation}
\hat{\Sigma}(i\omega_n) = \frac{Un}{2}\mathbf{1}-\frac{U m}{2}\sigma_z+\hat{\tilde{\Sigma}}(i\omega_n),\label{eq6}
\end{equation}
where the first two terms are the HF contribution, with $n=n_{\uparrow}+n_{\downarrow}$ and $m=n_{\uparrow}-n_{\downarrow}$, and local density $n_{\sigma}=\langle c^\dagger_{\sigma} c^{\phantom{\dagger}}_{\sigma} \rangle$ while the last term is the dynamical contribution of the local self-energy. We employ real-space dynamical mean-field theory (R-DMFT)~\cite{Snoek10} to calculate the local self-energy, combined with a continuous-time quantum Monte Carlo solver (CT-AUX)~\cite{Gull11}. 
  
 We use the above effective non-interacting  Hamiltonian $\hat{h}_t$ to calculate the $\mathbb{Z}_2$ invariant using the numerical method proposed by Fukui et al~\cite{Fukui07}, which uses twisted boundary conditions~\cite{Sheng06} and discretization of momentum. In the following, we briefly review their methodology. 
\subsection{Numerical calculation of the $\mathbb{Z}_2$ invariant using the effective topological Hamiltonian}
The method by Fukui et al~\cite{Fukui07} has similarities with the one proposed by Resta~\cite{Resta97} to  numerically calculate the polarisation of crystalline dielectrics, which is defined as the Berry phase of the electronic wave-function. We impose spin-dependent twisted boundary conditions along the $x$-direction, i.e. $c_{x+N_p,y}=e^{ik_p \sigma^z}c_{x,y}$ , and periodic boundary conditions along the $y$-direction, i.e. $c_{x,y+N_q}=c_{x,y}$.  We define the discretized quasi-momentum $\k_l\equiv(k_p, k_q)$ where $k_p=2\pi p/N_p$ and $k_q=2\pi q/N_q$ with $p\in \lbrace 0,....,N_p -1 \rbrace$ and $q\in \lbrace 0,...,N_q-1 \rbrace$. Further, we define the $U(1)$ link variable in the two quasi-momentum directions using the multiplet of the occupied states $\phi=\left( |a\rangle, |b\rangle,... \right)$ below the Fermi-level $E_F$, as
\begin{equation}
U_{1(2)}(\k_l)=\frac{\text{det} g_{1(2)} (\k_l)}{|\text{det} g_{1(2)} (\k_l)|}. \label{eq12}
\end{equation}
Here $g_{1(2)}$ are matrices with dimension equal to the number of states below the Fermi-level and the different  matrix elements can be given in terms of eigenstates $|a\rangle, |b\rangle \in \phi$ as
\begin{equation}
[g_{1}(\k_l)]_{ab} = \langle a(\k_l)|b(\k_l+\hat{\bf p})\rangle,
\end{equation}
\begin{equation}
[g_{2}(\k_l)]_{ab} = \langle a(\k_l)|b(\k_l+\hat{\bf q})\rangle,
\end{equation}
where $\hat{\bf p}$ and $\hat{\bf q}$ are unit vectors in the respective directions. Using the above link variable, one can define the Berry connection as
\begin{equation}
A_{1(2)}(\k_l)=\log{U_{1(2)}(\k_l)},\label{eq13}
\end{equation}
and define the field strength as 
\begin{equation}
F_{12}(\k_l)=\Delta_1 A_{2}(\k_l)-\Delta_2 A_1(\k_l),\label{eq14}
\end{equation}
where $\Delta_{1(2)}$ is the forward difference operator on the lattice given as $\Delta_{1(2)}=f(\k_l+\hat{\bf p}(\hat{\bf q}))-f(\k_l)$. Using Eq.(\ref{eq13}) and  Eq.(\ref{eq14}), the expression for Berry field strength can be written as
 \begin{equation}
 F_{12}(\k_l)=\log\left[{U_1(\k_l)U_2(\k_l+\hat{\bf p})U_1(\k_l+\hat{\bf q})^{-1}U_2(\k_l)^{-1}}\right].\label{15}
 \end{equation}
 
Finally, summing over all discrete $\k_l$, the Chern number is given as
 \begin{equation}
 \nu \equiv \frac{1}{4\pi i}\sum_l F_{12}(\k_l).\label{eq16}
 \end{equation}
 In the context of the QSH insulator, the topological invariant $\nu$ is equivalent to the $\mathbb{Z}_2$ invariant or spin-Chern number. The topological invariant obtained by this method correctly produces the continuum results even for a coarsely discretized Brillouin zone. It should be stressed that the
above calculations can be performed in any gauge. We do not need a specific gauge-fixing method to make the gauge connection smooth as done in the work by Fu and Kane~\cite{Fu07}. The 
 eigenvectors calculated using any numerical library can be employed to compute the topological invariant. Additionally, the method can naturally be used for inversion symmetry-broken cases, for example the TRI HH model with staggered field.
 The following is our algorithm for computing the topological invariant in the presence of interactions.
\begin{enumerate}
\item Calculate the self-energy for the TRI HH model using  DMFT for a set of model parameters.
\item Extract the self-energy at zero frequency after interpolating it to zero frequency and the self-energy at large frequency where it becomes constant (this is the HF part of the self-energy).
\item  Define the effective topological Hamiltonian according to Eq.(\ref{eq4}).
\item Diagonalize the effective Hamiltonian to obtain eigenvalues and eigenvectors. This is carried out using discretized momentum with twisted boundary conditions in $x$-direction.
\item For given filling (chemical potential), we calculate the spin Chern number $\nu$ (the $\mathbb{Z}_2$ topological invariant) using the scheme mentioned above. It is obtained from the eigenvectors of the effective non-interacting topological Hamiltonian.
\item We analyse the quantity $|\Sigma_R(i\omega_n \rightarrow 0)-\Sigma_{HF}|$ to comment on the role of dynamical contributions of the interaction when determining the QSH state. A large difference will signify that the topological state is mainly governed by dynamical effects of two-body interactions.  
\end{enumerate}
\subsection{Relation to ionic Hubbard model}\label{ionic}
\begin{figure}
\centering
\includegraphics[scale=0.32]{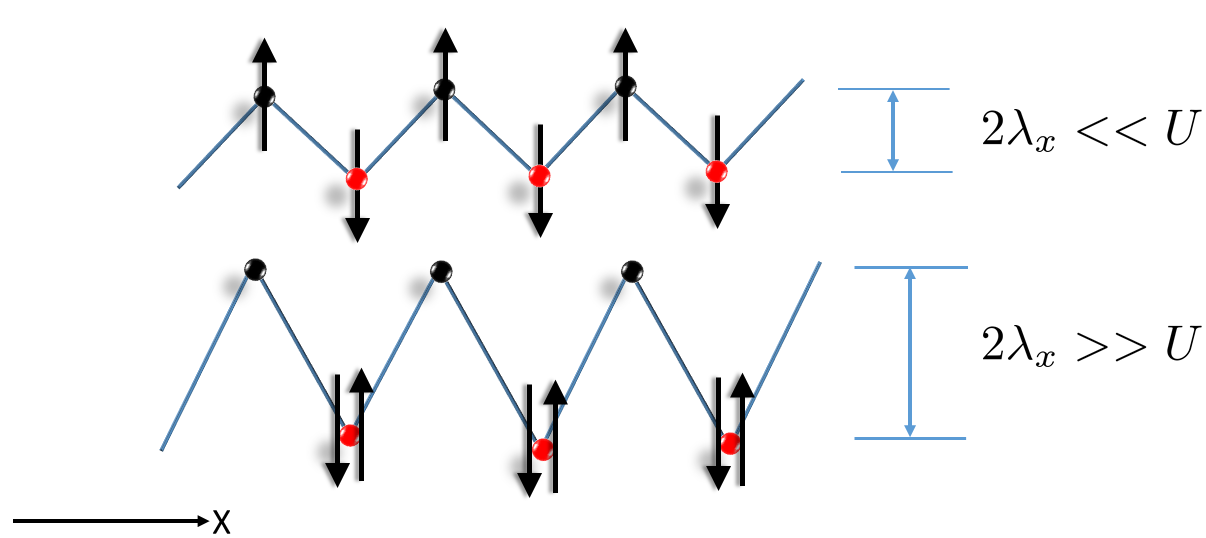}
\caption{Schematic representation of the possible electronic configurations in the absence of spin-orbit coupling. For large interactions ($U>>\lambda_x$), a magnetically ordered phase is energetically favoured while, in the presence of large staggering $U<<\lambda_x$  a band insulating phase with double occupation on alternating sites will be stable.}\label{fig1}
\end{figure}

The TRI HH model is related to the ionic Hubbard model in two dimensions which has been studied  previously~\cite{Bouadim07}. The model Hamiltonian presented here includes complex hopping (staggering is present only in $x$-direction). The schematic representation of the electronic configurations in the absence of spin-orbit coupling for the two limiting cases, i.e. (i) an on-site Coulomb interaction larger than the the staggering potential and (ii) the staggering potential dominating the on-site Coulomb interaction, is shown in Fig.~\ref{fig1}.
We define the staggered magnetization as
\begin{equation}
m_s=|m_A -m_B|, \label{eq19}
\end{equation}
where $m_{A/B}=n_{A/B\uparrow}-n_{A/B\downarrow}$ with $A\equiv\lambda_x$ and $B\equiv -\lambda_x$. Similarly the staggered filling can be defined as
\begin{equation}
 n_s=|n_A-n_B|,\label{eq20}
\end{equation}
where $n_{A/B}=n_{A/B\uparrow}+n_{A/B\downarrow}$. We also calculate the double occupancy $D=\langle n_{\uparrow}n_\downarrow \rangle$ to identify the transition from singly occupied magnetically ordered phase to doubly occupied band insulating phase. 

\section{Results and discussion}
As mentioned in the introduction, our main objective in this work is to elucidate the emergence and stability of the QSH state in the presence of two-body interactions. We present the complete phase diagram with and without spin-mixing. Throughout this article, we consider the half-filled case, i.e. the average number of particles per site is one. The inverse temperature has been chosen as $\beta=1/T=20$. We present the self-energy for different regions of the phase diagram and comment on the dynamical contribution of the self-energy on the QSH state. We show that in some parameter regimes the emergence of the QSH state is mainly governed by many-body interactions and can not be captured by a constant shift of model parameters in the non-interacting Hamiltonian. The QSH state is destroyed due to emergence of magnetic ordering (broken $SU(2)$ symmetry). We comment on the nature of the insulating state in the magnetically ordered region. We also compare our results with previous work~\cite{Cocks12,Orth13}, which classified topological states by counting of edge states in the bulk gap. Further, relating the present model to the ionic Hubbard model, we discuss staggered magnetization $m_s$, staggered occupancy $n_s$ and double occupancy $D=\langle n_\uparrow n_\downarrow \rangle$ as a function of the staggering potential $\lambda_x$ for given interaction strength $U$. 
\subsection{Interaction-driven QSH state for finite spin-mixing}
Without interactions, i.e. for $U=0$, the TRI HH model shows a transition from QSH state to the trivial BI~\cite{Cocks12} with increasing staggering field $\lambda_x$ for fixed spin-mixing $\gamma=0.25$. We choose specifically $\gamma=0.25$ for our calculations, since here the QSH state persists in a wide region of the phase diagram when $U=0$.
\begin{figure}[h!]
\centering
\includegraphics[scale=0.28]{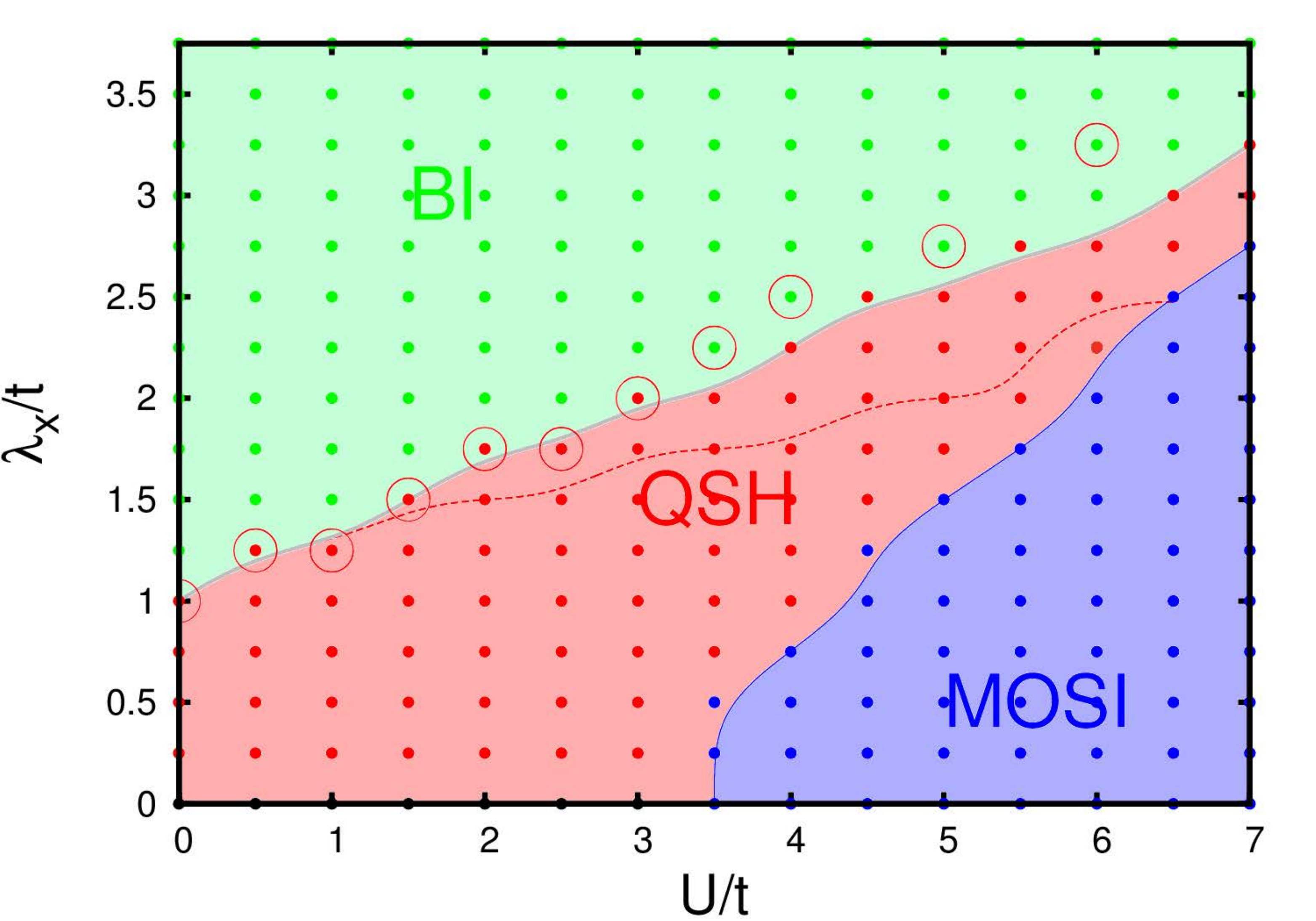}
\caption{Phase diagram of the TRI HH model for $U$, $\lambda_x$, $\gamma=0.25$ and $\alpha=1/6$ obtained using the topological Hamiltonian combined with DMFT for half-filling and inverse temperature $\beta=1/T=20$. Solid lines are obtained using the full DMFT local self-energy while the dashed line is the phase boundary between BI and QSH state extracted using the HF (static) part of the self-energy ($\Sigma(i\om_n\rightarrow \infty)$). For the non-interacting case ($U=0$), the system is a trivial BI for $\lambda_x > 1.25$ and non-trivial QSH for $\lambda_x  < 1.25$. For large $U$, the system is a magnetically ordered Slater insulator (MOSI). The QSH state can be induced by tuning the interaction strength $U$ from small to moderate values for finite $\lambda_x$. The encircled points are evaluated by counting the number of Kramers-pairs of helical edge modes in the gap.  }\label{fig2}
\end{figure}
We calculate the phase diagram as shown in Fig.~\ref{fig2} for different values of interaction $U$ and staggering field $\lambda_x$. It is clear that for $\lambda_x > 1.25$, the system traverses three different phases as the interaction strength is increased:  $1)$ In the weakly interacting regime, the system is in the topologically trivial BI regime (green dots). $2)$ At intermediate values of the interaction strength $U$, the system is in the QSH state (red dots). $3)$ For further increasing $U$ the system undergoes a transition to a magnetically ordered Slater insulating (MOSI) state (blue dots). The magnetic state has collinear ordering as reported previously~\cite{Cocks12,Orth13}. The solid line is obtained from the full DMFT self-energy at zero frequency ($\Sigma(i\om_n\rightarrow 0)$) and the dashed line is the phase boundary between QSH and BI extracted from the HF self-energy ($\Sigma(i\om_n\rightarrow \infty)$).  The HF part of the self-energy, for $m=0$ as mentioned before, is $Un/2$ ($n=n_\uparrow + n_\downarrow$) where the local density $n_\sigma$ can be extracted from the local Green's function i.e. $n_\sigma=G_\sigma(\tau\rightarrow 0^{-})$. For large values of interaction and staggering, the topological QSH region of the phase diagram gets narrow and there will be a phase transition directly from BI to the magnetically ordered phase. Here, it is important to note that the gap closes along the transition line from topologically trivial BI to QSH state. Also, for $\lambda_x=0$ the system is SM below the critical $U$ above which $m$ gets finite. The phase diagram is the result of competition of the band-insulator gap due to applied staggering potential with the local interaction $U$~\cite{Werner07,Budich13,Amaricci16}. As mentioned in the introduction, we additionally evaluate the phase boundary between the QSH and BI by determining the topological index by counting the number of Kramers-pairs of helical edge modes crossing the bulk gap~\cite{Cocks12,Orth13}. An even number of pairs (including zero) of helical edge modes corresponds to the trivial band insulator, while an odd number of pairs corresponds to the QSH state. The result is represented by the encircled points in the phase diagram. In order to elucidate the method, in Fig.~\ref{fig3} we show the intensity plot of the spectral function for particular values of $U$ and $\lambda_x$ in the QSH and BI phases. For this purpose, real space DMFT calculations were carried out to calculate the spectral function $A(x, k_y, \om)$ where $A(x, k_y, \om)=-\frac{1}{\pi} \Im G(x, k_y, \om)$. We use open boundary conditions in $x$-direction and periodic boundary conditions in $y$-direction. We have used the maximum entropy method to extract the real-frequency Green's function~\cite{Jarrell96, Wang09}.
\begin{figure}[h!] 
\centering
\includegraphics[scale=0.42]{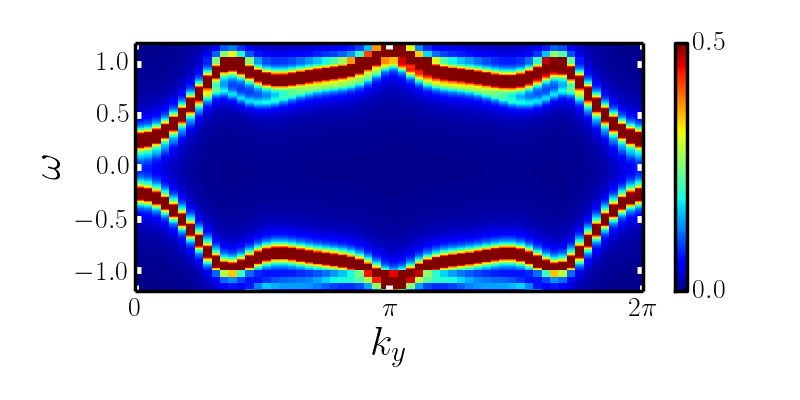}
\includegraphics[scale=0.42]{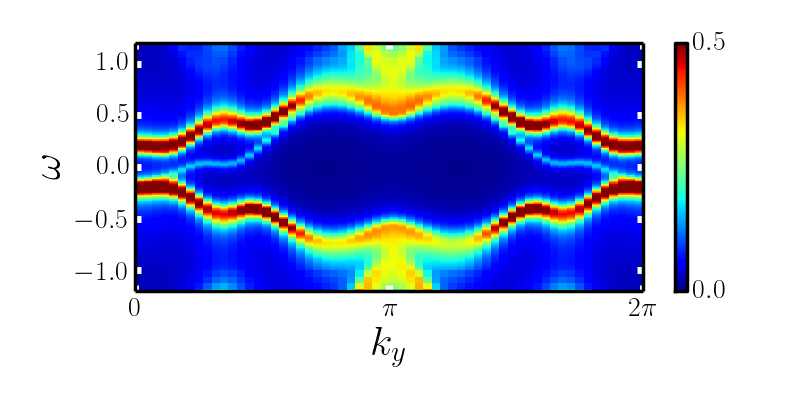}
\caption{Upper panel: Intensity plot of the single-particle spectral function $A(k_y ,\omega)$ for $\lambda_x=2$ and $U=1$. Lower panel: Intensity plot of the single-particle spectral function for $\lambda_x=2$ and $U=4$; the presence of a single pair of edge modes crossing the bulk gap and connecting the two bulk bands confirms the QSH state. The spin mixing parameter is $\gamma=0.25$.}\label{fig3}
\end{figure}
In the upper panel of Fig.~\ref{fig3}, we show the intensity plot of the single-particle spectral function $A(k_y,\om)=\sum_x A(x, k_y, \om)$ for a set of parameters in the band insulating regime of the phase diagram in Fig.~\ref{fig2}. We observe that for these parameters ($\lambda_x=2$, $U=1$ and $\gamma=0.25$), there is no edge state present in the bulk gap and hence the system is a topologically trivial BI. In the lower panel of Fig.~\ref{fig2} we choose parameters ($\lambda_x=2$, $U=4$ and $\gamma=0.25$) from the QSH region of the phase diagram of Fig.~\ref{fig2}. There is now a single pair of helical edge modes traversing the bulk gap and thus the system is in the QSH state. This analysis gives an example of the equivalence of the topological Hamiltonian and the bulk-boundary correspondence for the topological index. It is important to note here that the choice of parameters is such that the dynamical contribution of the many-body interaction gets significant. In accordance with the previous literature~\cite{Hohenadler13}, we also find that up to moderate values of interaction ($U \approx 3$) both methods give same quantitative results. The two methods deviate further for large $U$. The deviation could be due to finite  size of the stripe geometry. Additionally, the extraction of the self-energy at zero frequency is carried out after numerical interpolation. This could lead to further numerical error.
 To understand the dynamical contribution of the self-energy, we show $\Sigma_R(i\omega_n)$ for $\lambda_x=2.25$ and $\lambda_x=1$ for different $U$ values in Fig.~\ref{fig4} in the upper and lower panel respectively.
\subsection{Local self-energy}
\begin{figure}[h!]
\centering
\includegraphics[scale=0.4]{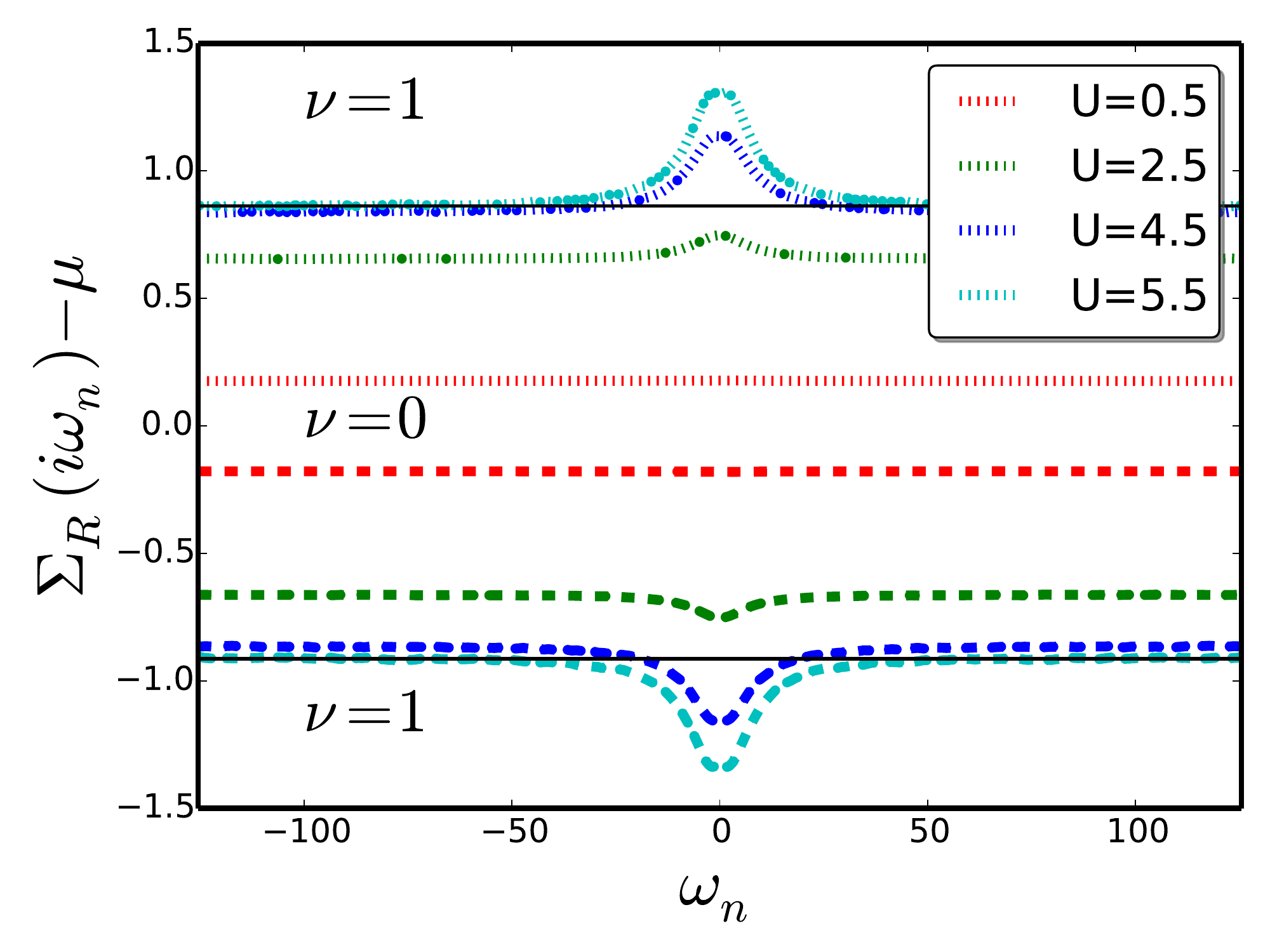}
\includegraphics[scale=0.4]{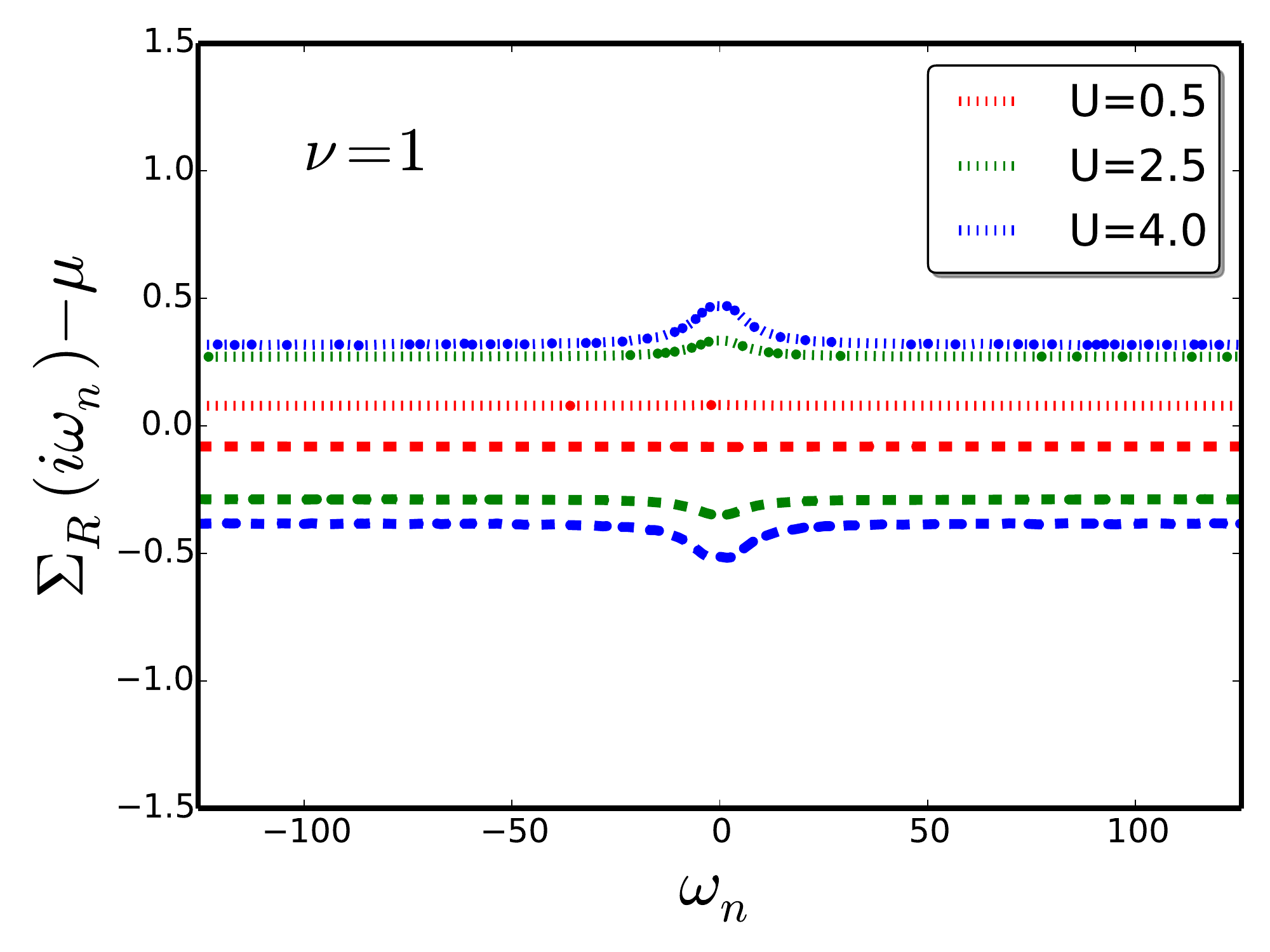}
\caption{Upper panel: Real part of the self-energy for $\lambda_x=2.25$ for different $U$. If the value of $\Sigma_R(i\omega_n)-\mu$ is between the solid lines at all $\omega_n$, we find the system to be a trivial insulator; otherwise it is in the QSH state. Lower panel: Real part of the self-energy for $\lambda_x=1$. The system is in a QSH state with finite topological $\mathbb{Z}_2$ invariant i.e. $\nu=1$. The dotted lines correspond to staggering field $\lambda_x$ while the dashed lines are for staggering field $-\lambda_x$.}\label{fig4}
\end{figure}

As discussed in section~\ref{subsec:formalism}, the effective topological Hamiltonian depends upon the zero frequency interpolated self-energy. We choose two values of the staggering field, $\lambda_x=2.25$ (upper panel) and $\lambda_x=1$ (lower panel), and show the self-energies for different values of $U$ in Fig.~\ref{fig4}. In the upper panel, the real part of the self-energy, i.e $\Sigma_{\text{R}}(i\omega_n)$, is constant versus frequency for small values of $U$ (in the BI regime) and thus $\Delta \Sigma=|\Sigma_R(i\omega_n \rightarrow 0)-\Sigma_{HF}|\simeq 0$ where $\Sigma_{HF}=\Sigma(i\omega_n \rightarrow \pm \infty)$. For intermediate values of $U$ (in the QSH state) we have $\Delta \Sigma > 0$ and thus the dynamical contribution of the self-energy becomes significant.  Further, with increasing value of $U$, the quantity $\Delta \Sigma$ gets more prominent but still remains finite in the magnetically ordered regime. 
In the lower panel, we perform a similar analysis for $\lambda_x=1$, when the system goes through a transition from the QSH state (for moderate values of $U$) to the MOSI state (for large $U$), and the quantity $\Delta \Sigma \simeq 0$ for different interaction strengths. From the above analysis of the self-energy in various regions of the phase diagram, we can conclude that for moderate values of the staggering field, HF theory captures the essential details of the topological QSH insulator in the presence of finite interactions. For large staggering field and large interaction strength, when the dynamical part of the self-energy gets prominent, it is necessary to carry out calculations beyond static HF theory, for example by DMFT, to capture the dynamical effect of many-body interactions.
 
\subsection{Magnetic phase boundary}
\begin{figure}[h!]
\begin{center}
\includegraphics[scale=0.45]{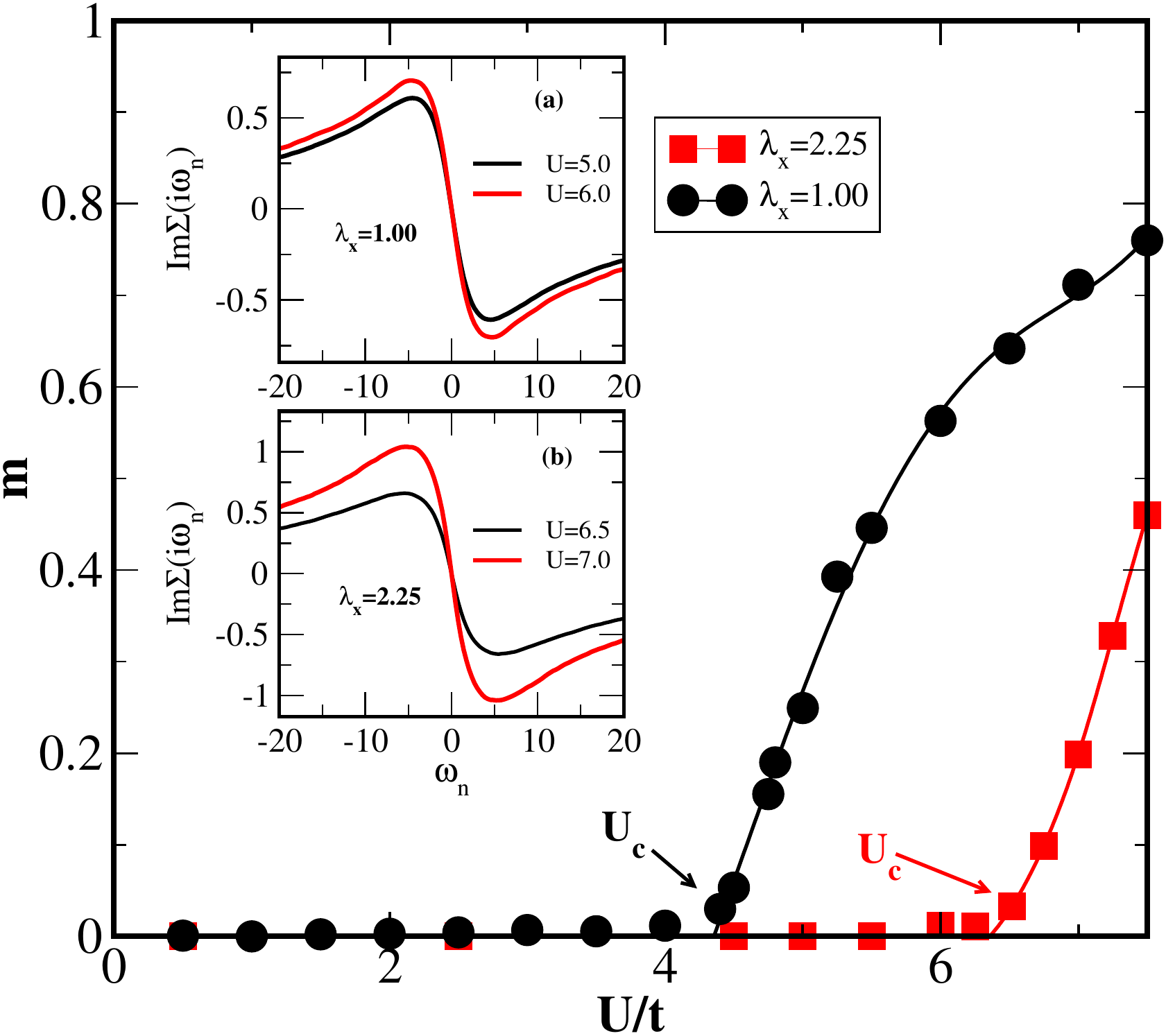}
\caption{Local magnetization for different values of $U$ for $\lambda_x=1$ and $2.25$. $U_c$ is the critical value of $U$ above which the local magnetization is finite and SU($2$) symmetry broken. Circles (squares) are the actual data points and solid lines are a guide to the eye. We show the imaginary part of the self-energy, for $\lambda_x=1$ in the inset (a) and for $\lambda_x=2.25$ in the inset (b), for interaction strengths $U$ for which $m > 0$.}\label{fig5}
\end{center}
\end{figure}

In the phase diagram shown in Fig.~\ref{fig2}, with increasing interaction strength $U$, there emerges a finite local magnetization ($m= n_{\uparrow}-n_{\downarrow}$) and thus the $SU(2)$ symmetry breaks down and the QSH state is destroyed. In the main panel of Fig.~\ref{fig5}, we show the local magnetization $m$ versus $U$ for $\lambda_x=2.25$ and $\lambda_x=1$.  For a given value of $\lambda_x$ there is a threshold value $U_c$ above which the local magnetization assumes a finite value. With increase of the staggered potential, the critical value $U_c$ also increases.
Magnetic phases for large interaction strength have been obtained in previous work~\cite{Cocks12,Orth13}, but the insulating nature of these magnetically ordered states has not been discussed in detail. In order to understand the nature of the gap in the magnetically ordered region, we present the imaginary part of the local self-energy in the inset of Fig.~\ref{fig5} for different interaction strengths when $m > 0$, for (a) $\lambda_x=1$ and (b) $\lambda_x=2.25$. The imaginary part of self-energy has a linear dependence for small frequencies, i.e. $\Im \Sigma(i\om_n)\propto i\omega_n$, and thus the band picture still holds true. With increasing value of $U$, $\Im \Sigma(i\omega_n)$ increases in the vicinity of zero frequency~\cite{Sentef}. The band insulating magnetism shown in Fig.~\ref{fig4} has been termed Slater insulator (SI) and has been explored previously in the context of the Hubbard model~\cite{Pruschke03} using DMFT.  
\begin{figure}[!ht]
\includegraphics[width=0.4\textwidth,height=50mm]{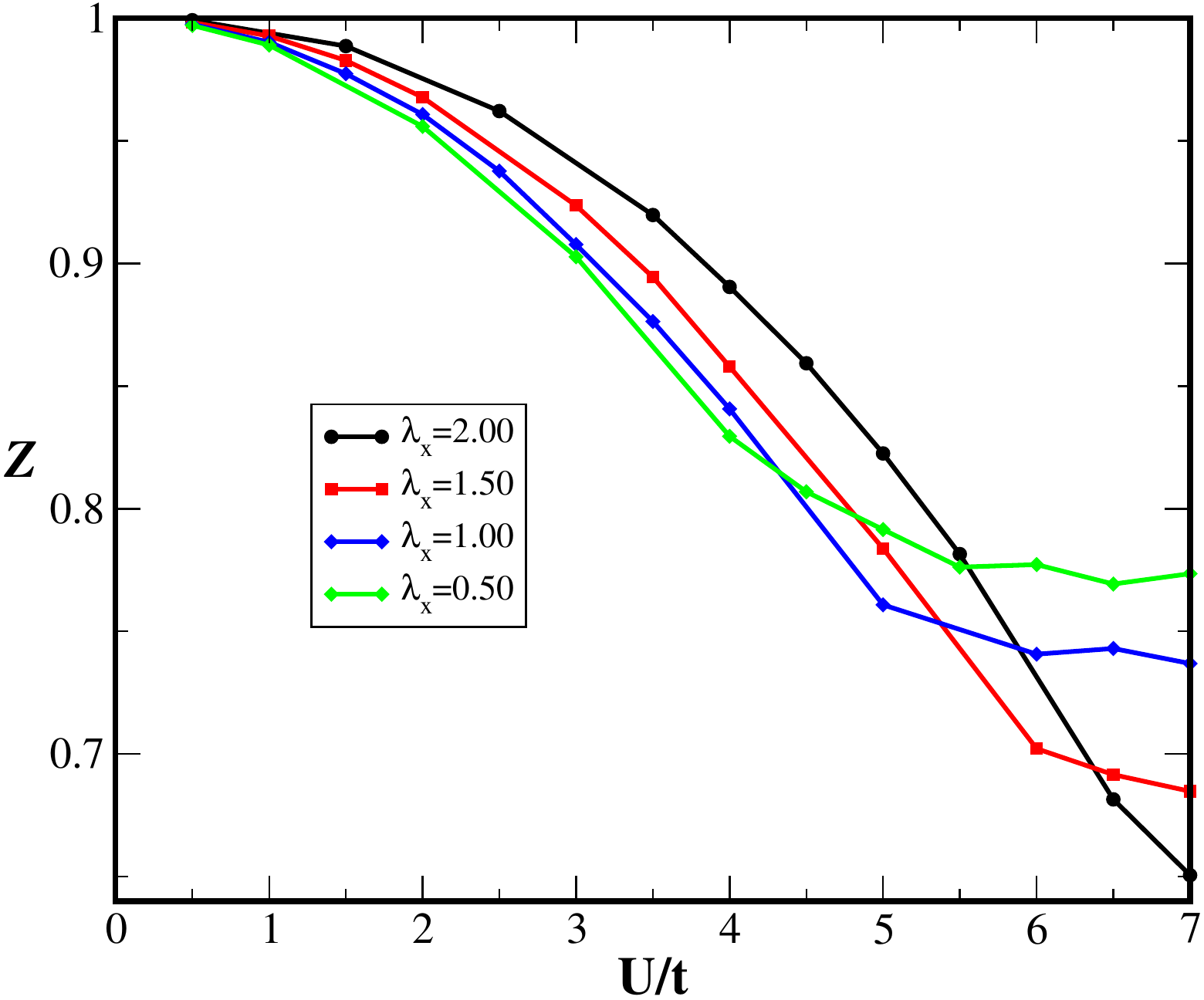}
\caption{Quasi-particle weight $Z=\left(1-\frac{\partial \tiny{\Im} \Sigma (i\omega))}{\partial \omega_n}|_{\omega_n\rightarrow 0}\right)^{-1}$ for varying interaction strengths and different values of the staggering potential $\lambda_x$.}\label{fig6}
\end{figure}  
 In Fig.~\ref{fig6}, we show the quasi-particle weight versus the interaction strength $U$ for different values of the staggering potential $\lambda_x$. The quasi-particle weight decreases with increasing $U$, but has a well-defined finite contribution even in the magnetically ordered regime, which signifies that the system has BI behaviour. It is important to note that the change of quasi-particle weight as a function $U$ is slower in the magnetically ordered phase than in the non-magnetic region.
\subsection{Phase diagram for zero spin-mixing}
\begin{figure}[ht]
\centering
\includegraphics[scale=0.28]{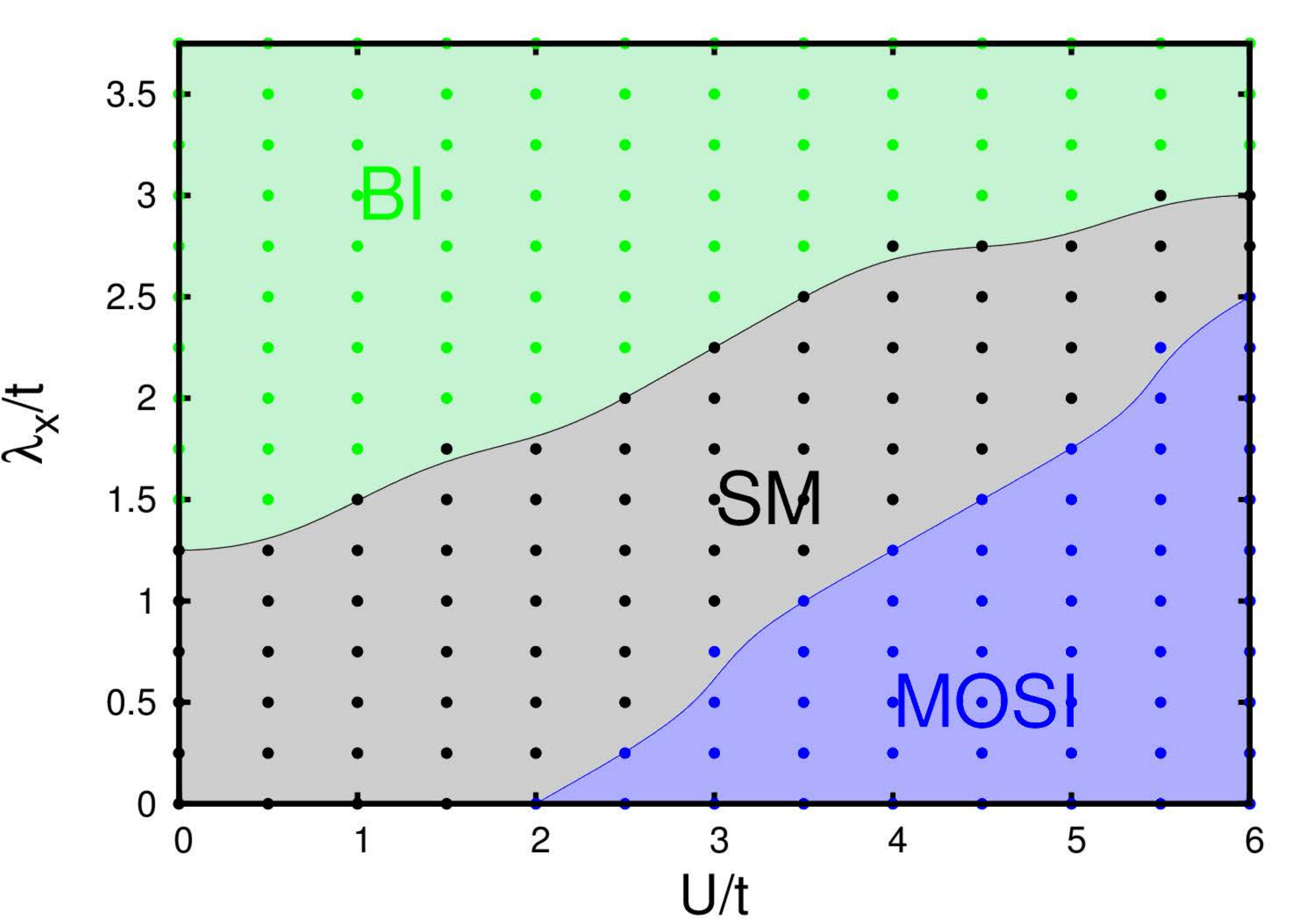}
\caption{Phase diagram of the TRI Hofstadter Hubbard model for $\alpha=1/6$ and $\gamma=0$ obtained by the topological Hamiltonian in combination with DMFT. For moderate values of interaction strength and staggering, the system is in the semi-metallic phase. No QSH state emerges here.}\label{fig7}
\end{figure}

We also employ the topological Hamiltonian for zero spin-mixing, i.e. $\gamma=0$, and present the phase diagram for varying interaction $U$ and staggering field $\lambda_x$ in Fig.~\ref{fig7}. 
For staggering potential $\lambda_x > 1.25$, the system traverses three different phases as the interaction strength is increased: $1)$ In the weakly interacting regime, the system is in the topologically trivial BI regime (region shown with green dots). $2)$ At intermediate values of the interaction strength $U$, the system is a gapless SM (region shown with black dots). $3)$ With further increasing U the system undergoes a first order transition to the MOSI state. The magnetically ordered state  has anti-ferromagnetic (AF) N{\'e}el ordering and is different from the finite spin-mixed case, i.e $\gamma=0.25$, which shows collinear magnetic ordering~\cite{Cocks12, Orth13}. It is important to note that there is no QSH state present without spin-mixing, similar to the case of $U=0$~\cite{Cocks12}.
\subsection{Staggered magnetization}
\begin{figure}[h!]
\includegraphics[scale=0.55]{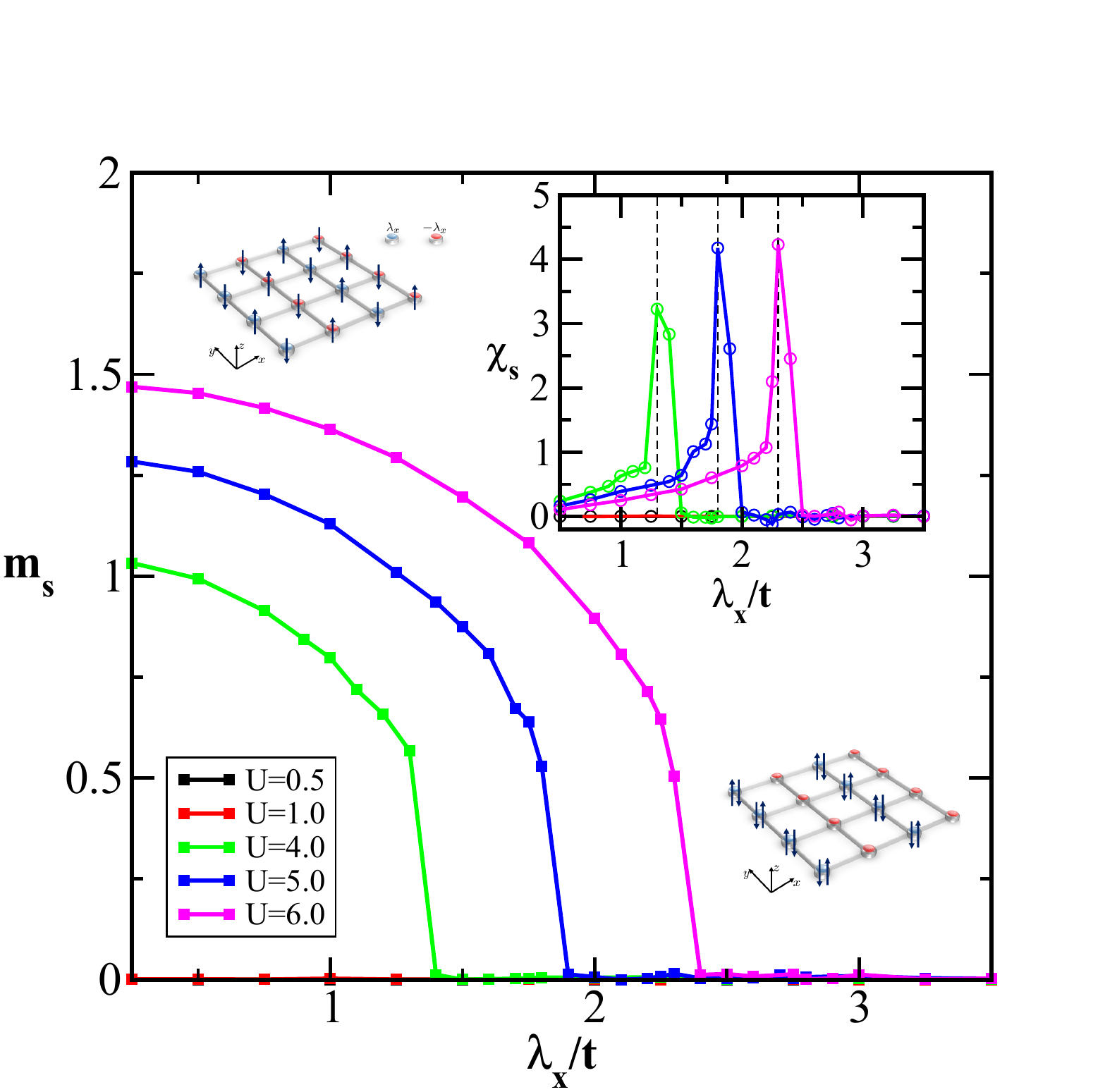}

\caption{Staggered magnetization $m_s$ as a function of staggering $\lambda_x$ for different interaction strengths $U$. The staggered magnetization is always zero below a critical $U_c$. There is a first order transition with $\lambda_x$ for $U>U_c$. In the inset, we show the susceptibility $\chi_s=-dm_s/d\lambda_x$ which is sharply peaked and discontinuous at a critical staggering $\lambda^c_x$ for $U>U_c$. The schematic and idealized representations of magnetic and particle density ordering are  shown for finite and zero staggered magnetization.}\label{fig8}
\end{figure}

The magnetic ground state for different values of the spin-mixing has been discussed in previous work~\cite{Cocks12,Orth13}. The system goes from an AF N{\`e}el ordered state at $\gamma=0$ through spiral ordering at intermediate values of the spin-mixing to collinear magnetic ordering at $\gamma=0.25$. The effect of staggering $\lambda_x$ on these magnetically ordered states has not been explored yet.  The staggered magnetization $m_s$, defined in section~\ref{ionic}, will always vanish for collinear magnetic ordering with ferromagnetic ordering in $x$-direction and anti-ferromagnetic ordering in $y$-direction for $\gamma=0.25$~\cite{Cocks12, Orth13}. We choose $\gamma=0$ to explore the nature of the magnetic transition with staggered potential. In the main panel of Fig.~\ref{fig8}, we show the staggered magnetization $m_s$ for varying staggering $\lambda_x$ at different interaction strengths $U$. For a given value of $U$ there exists  a critical value of the staggering  $\lambda_x^c$, at which the magnetization vanishes with a first order transition. The magnetic instability does not occur at arbitrarily small values of $U$ for $\lambda_x=0$, instead a critical value $U^c$ is required to induce staggered magnetization. Both $m_s$ at $\lambda_x=0$ and the critical value of the staggering potential, $\lambda_x^c$, are approximately proportional to the interaction strength $U$. In the inset the susceptibility, defined as $\chi_s=-dm_s/d\lambda_x$ is shown versus the staggering potential. It has a sharply peaked behavior with a discontinuity  at the transition point.
\subsection{Occupancy}
In this section, we discuss the effect of the staggered potential $\lambda_x$ on the staggered occupancy $n_s$ and the double occupancy $D=\langle n_{\uparrow} n_{\downarrow}\rangle$ for different interaction strengths. 
\begin{figure} [h!]
\includegraphics[scale=0.39]{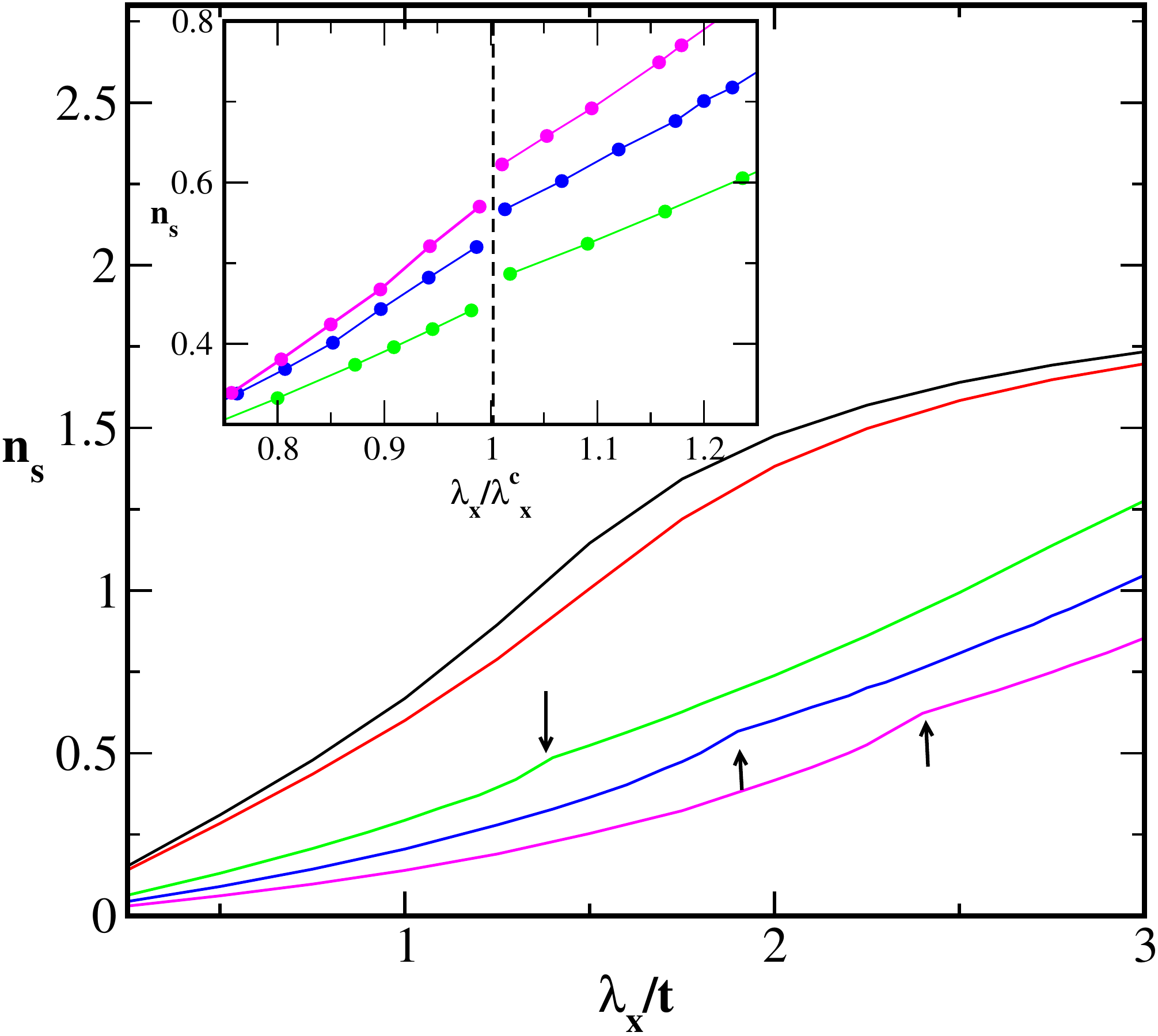}
\includegraphics[scale=0.35]{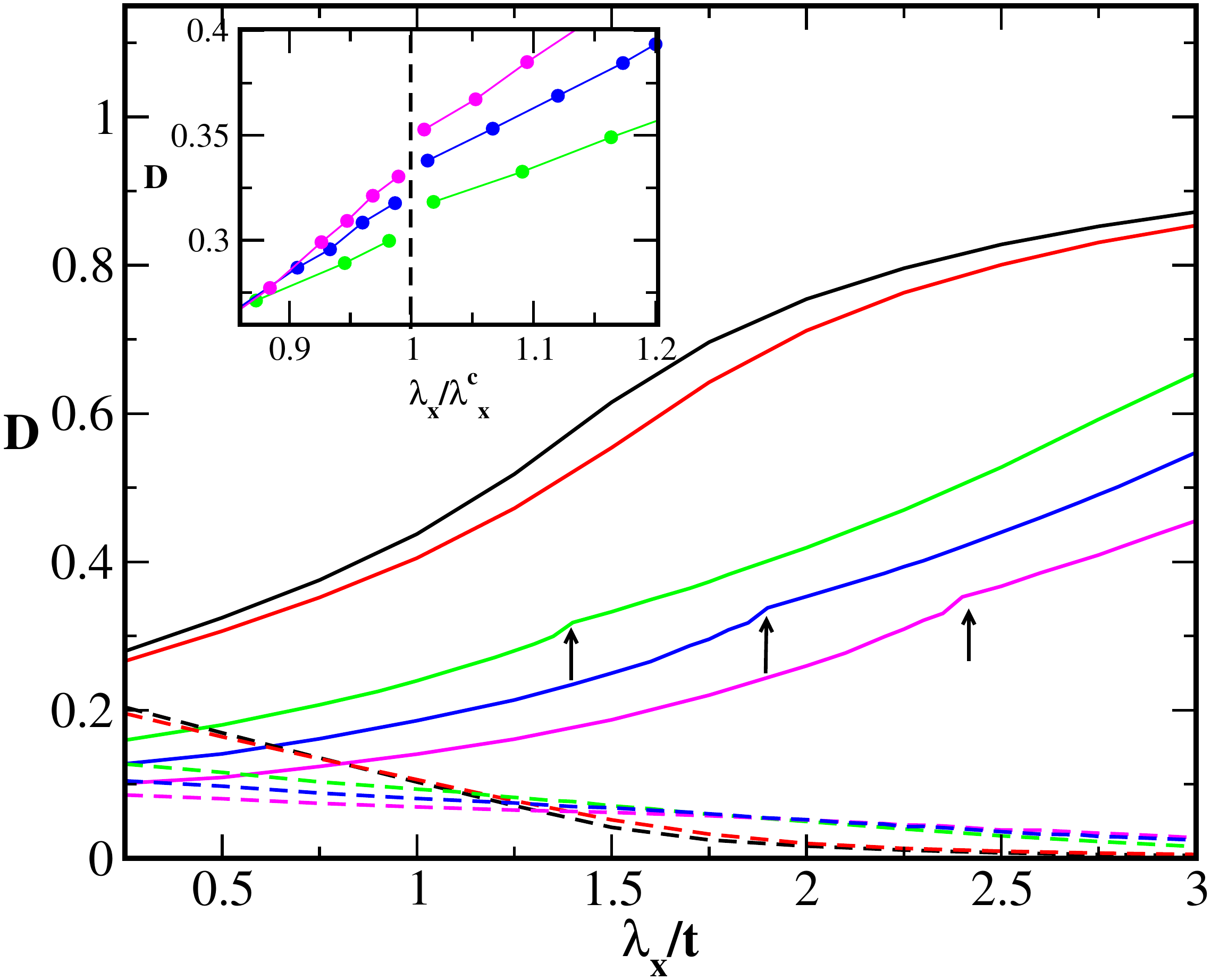}
\caption{Upper panel: We show the staggered occupancy $n_s$ as a function of staggered potential $\lambda_x$ for different interaction strengths $U$. There is a first order transition at $\lambda^c_x$ for $U>U_c$ (shown by arrows). In the inset, $n_s$ versus scaled staggering potential with corresponding $\lambda^c_x$ for $U>U_c$ is shown. The discontinuity is clearly visible. Lower panel: We show the double occupancy $D$ versus staggering field for different interaction strengths $U$ (transition points shown by arrows). The solid (dashed) line denotes sites with onsite potential $\lambda_x$ ($-\lambda_x$). In the inset, the discontinuity for $U>U_c$ is apparent. Color code is according to figure~\ref{fig8}.\label{fig9}}
\end{figure}
In the main panel of upper panel of Fig.~\ref{fig9}, we show the staggered occupancy $n_s$, as defined in section~\ref{ionic}. For $\lambda_x=0$, $n_s$ vanishes and increases monotonically with $\lambda_x$. With increasing interaction strength, $n_s$ decreases at fixed $\lambda_x$. A discontinuity appears exactly at the critical value $\lambda_x/\lambda_x^c\approx 1$, for $U>U_c$ shown in the inset. In the main panel of lower Fig.~\ref{fig9}, we show the double occupancy $D$ at two sites with staggered potential $\lambda_x$ and $-\lambda_x$. In the absence of staggering, the double occupancy at both sites is equal and decreases with increasing U (for large $U$ in the Mott regime $D$ goes to zero). With increasing $\lambda_x$ a potential difference $2\lambda_x$ between the two sites arises and thus one site prefers double occupancy while the other site is empty. The Hubbard interaction $U$ competes with the staggering $\lambda_x$ in order to suppress the double occupancy. Also in this case a discontinuity appears at $\lambda_x^c$ as shown in the inset. Such a discontinuous transition visible in the order parameters has been reported for ionic Hubbard model with similar method~\cite{Bag15}.\\
\section{Summary and outlook}
We have applied the effective topological Hamiltonian approach in combination with HF and DMFT to capture interaction effects on topological bands. We have explored the TRI HH model and shown the presence of the QSH state, which is in selected parameter regimes governed by the dynamical contribution of the two-body interactions. We have discussed the effect of the static and dynamical parts of the self-energy on the emergence of the QSH state. The magnetic phase observed is a SI and this has been argued clearly based on the nature of the self-energy. The magnetically ordered region has a well-defined finite quasi-particle weight, thus ensuring that the nature of the gap is BI. We have  compared  our findings with the edge mode spectrum and have confirmed the bulk-boundary correspondence. The TRI HH model explored here is similar to a $2$D ionic Hubbard model with complex hopping. The staggered magnetization shows a first order transition versus staggered potential at finite interaction.  We also discuss the behaviour of staggered occupancy and double occupancy. Recent experimental findings for the ionic Hubbard model~\cite{Messer15} on the honeycomb lattice supports the idea of  extending it to a square lattice in combination with complex hopping. In the present work, we only consider the local self-energy approximation, spatial fluctuations are ignored. Methods such as cellular DMFT~\cite{Chen15} combined with the topological Hamiltonian can be used to explore the effect of these  spatial fluctuations on the QSH state and will be the subject of future investigation.
\acknowledgments
{ Support by the German Science Foundation Deutsche Forschungsgemeinschaft (DFG) via Sonderforschungsbereich Grant No.\ SFB/TR 49, Forschergruppe FOR 2414 and the high-performance computing
center LOEWE-CSC is gratefully acknowledged. The authors would like to thank Karyn Le Hur, Anand Sharma, Daniel Cocks, Andrii Sotnikov and Lei Wang for
fruitful discussions.
}

\end{document}